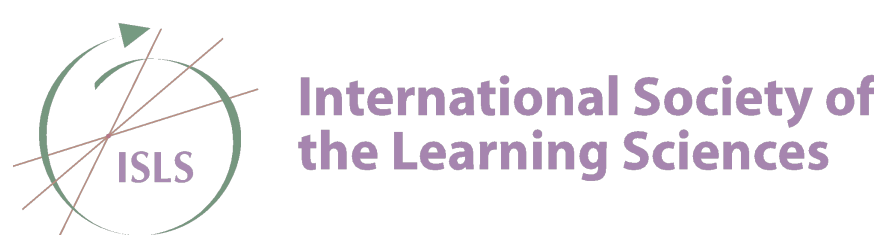

# How Motivation Relates to Generative AI Use: A Large-Scale Survey of Mexican High School Students

Echo Zexuan Pan, Harvard University, echozexuanpan@g.harvard.edu
Danny Glick, Oranim College of Education, danny.glick@oranim.ac.il
Ying Xu, Harvard University, yingxu@g.harvard.edu

**Abstract:** This study examined how high school students with different motivational profiles use generative AI tools in math and writing. Through *K*-means clustering analysis of survey data from 6,793 Mexican high school students, we identified three distinct motivational profiles based on self-concept and perceived subject value. Results revealed distinct domain-specific AI usage patterns across students with different motivational profiles. Our findings challenge one-size-fits-all AI integration approaches and advocate for motivationally-informed educational interventions.

## Introduction

Generative artificial intelligence (AI) has entered educational spaces at an unprecedented pace (Kasneci et al., 2023). These technologies hold considerable promise for providing personalized tutoring, explaining difficult concepts in multiple ways, and offering immediate feedback on student work, yet they have also sparked concerns about academic integrity, overreliance, and cognitive offloading—where students transfer cognitive effort to external tools rather than developing their own problem-solving skills (Giannakos et al., 2025; Lee et al., 2024; Ng et al., 2025). What underlies these mixed reactions is, in part, a fundamental gap in our understanding of how students actually use generative AI tools in their day-to-day learning.

What makes this issue even more complex is that students do not approach learning technologies as a homogeneous group. Decades of educational research have shown that students vary considerably in their academic motivation, and these differences shape how they engage with learning opportunities (Eccles & Wigfield, 2002; Wigfield & Eccles, 2000). According to situated expectancy-value theory, students' motivation stems from two key beliefs: their self-concept about own competence in a domain and their subjective values regarding that domain (Eccles & Wigfield, 2020). These motivational beliefs are not static traits but are situated within specific contexts and can vary across different subjects and situations.

The current study addresses this gap by examining how students with different motivational profiles use generative AI tools in their learning. By connecting motivational theory to emerging patterns of AI use, we aim to provide a more nuanced understanding of student experiences that can inform both educational practice and the design of AI-supported learning environments.

## Methods

### Participants

In collaboration with the Ministry of Education in Chihuahua (the second-largest state in Mexico) and public high schools across the state, we conducted an anonymous online survey with 7,739 high school students in their final two years (61.5% female). To ensure data quality and relevance, we excluded responses from students who failed attention checks or indicated no prior experience with generative AI tools, resulting in a final analytical sample of 6,793 students.

### Measures

Student motivation was assessed through two key constructs derived from situated expectancy-value theory (Eccles & Wigfield, 2020): self-concept (beliefs about one's own abilities) and perceived subject value (beliefs about the importance and enjoyment of a subject). Self-concept was measured using a 5-item survey, with each item rated on a scale ranging from 1 (*not good at all*) to 7 (*extremely good*), while perceived subject value was assessed using a 3-item survey, with each item rated on a scale ranging from 1 (*not at all*) to 7 (*very much*). Students reported their self-concept and perceived subject value across two subject domains—math and writing—under scenarios with and without generative AI availability.

AI usage patterns were assessed separately for each subject domain using a Likert scale from 1 (*never*) to 5 (*almost always*). For math, students reported their frequency of using AI for formula/background retrieval, direct answer copying, problem interpretation, step-by-step guides, solution verification, and practice problem generation. For writing, they reported their frequency of using AI for idea brainstorming, outline production, supporting evidence, language/grammar improvement, writing feedback, and requirement adaptation. All instruments were administered in Spanish and subsequently translated into English for analysis.

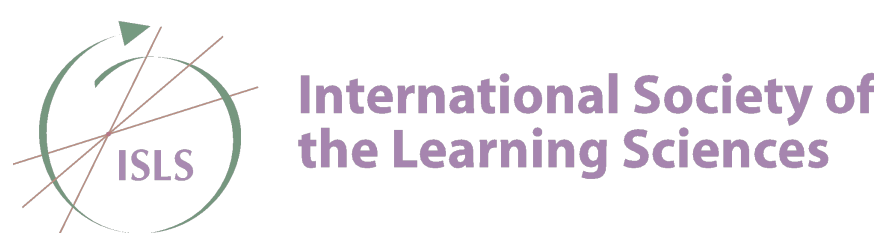

### Analysis

We conducted *K*-means clustering analysis based on students' self-concept and perceived subject value, performing separate analyses for the math and writing domains. The optimal number of clusters was determined through convergent evidence from three statistical criteria: the elbow method, silhouette analysis, and the gap statistic. All three methods consistently indicated that a three-cluster solution provided the most meaningful and interpretable partitioning of the data across both domains. Following cluster identification, we conducted pairwise comparisons to examine differences in each specific way of AI usage across the three motivational profiles.

## Results

### Three motivational profiles

The clustering analysis revealed three distinct motivational profiles that emerged consistently across both math and writing domains. The largest profile was characterized as *Aspirational*, representing students who highly valued the subject despite having low confidence in their own abilities. The second profile, *Confident*, represented the most academically engaged students who combined strong confidence in their abilities with high appreciation for the subject's value. The third profile, characterized as *Disengaged*, represented students who demonstrated both low confidence in their abilities and low appreciation for the subject's value.

### AI usage pattern by motivational profiles

In math, the three motivational profiles demonstrated distinct approaches to AI usage (Figure 1, upper). The Aspirational group ($n = 2,809$) used AI primarily as a compensatory tutor. These students reported the highest frequency of using AI for step-by-step problem guides ($M = 2.81$) and for interpreting or rephrasing complex problems ($M = 2.59$). The Confident group ($n = 2,207$) exhibited selective usage consistent with treating AI as a partner for refinement and exploration. They were less likely to rely on AI for fundamental tasks such as formula retrieval ($M = 2.50$) and step-by-step guides ($M = 2.60$) compared to the Aspirational group ($M = 2.64$, $p < .001$; $M = 2.81$, $p < .001$). Additionally, they showed distinctively higher usage of AI for generating practice problems ($M = 2.48$) compared to Disengaged students ($M = 2.32$, $p < .001$). The Disengaged group ($n = 1,777$) displayed a concerning pattern of shortcut-seeking behavior. While they used AI for formula retrieval ($M = 2.65$) and step-by-step guides ($M = 2.80$) at rates similar to Aspirational students, they reported the highest frequency of directly copying answers ($M = 2.69$), significantly exceeding both Aspirational ($M = 2.53$, $p < .001$) and Confident students ($M = 2.18$, $p < .001$).

In writing, analogous but distinct patterns emerged across the three motivational profiles (Figure 1, lower). The Aspirational group ($n = 3,191$) again demonstrated a compensatory tutoring pattern. They were most likely to use AI for brainstorming ideas ($M = 2.83$) and producing outlines ($M = 2.72$). The Confident group ($n = 1,915$) used AI primarily as a partner for refinement and quality assurance. They reported the highest frequency of using AI for language and grammar improvement ($M = 2.79$), as well as for seeking writing feedback ($M = 2.67$). The Disengaged group ($n = 1,687$) reported the lowest overall AI usage across most writing tasks.

## Conclusion

Our results showed that students with different motivational profiles used AI in different ways, and these usage patterns are domain-specific. These findings challenge one-size-fits-all approaches to AI integration and advocate for more targeted, motivationally-informed interventions that recognize students' characteristics.

## References

Eccles, J. S., & Wigfield, A. (2002). Motivational beliefs, values, and goals. *Annual Review of Psychology*, *53*(1), 109–132.

Eccles, J. S., & Wigfield, A. (2020). From expectancy-value theory to situated expectancy-value theory: A developmental, social cognitive, and sociocultural perspective on motivation. *Contemporary Educational Psychology*, *61*, 101859.

Giannakos, M., Azevedo, R., Brusilovsky, P., Cukurova, M., Dimitriadis, Y., Hernandez-Leo, D., Järvelä, S., Mavrikis, M., & Rienties, B. (2025). The promise and challenges of generative AI in education. *Behaviour & Information Technology*, *44*(11), 2518–2544.

Kasneci, E., Sessler, K., Küchemann, S., Bannert, M., Dementieva, D., Fischer, F., Gasser, U., Groh, G., Günnemann, S., Hüllermeier, E., Krusche, S., Kutyniok, G., Michaeli, T., Nerdel, C., Pfeffer, J., Poquet, O., Sailer, M., Schmidt, A., Seidel, T., … Kasneci, G. (2023). ChatGPT for good? On opportunities and challenges of large language models for education. *Learning and Individual Differences*, *103*, 102274.

Lee, V. R., Pope, D., Miles, S., & Zárate, R. C. (2024). Cheating in the age of generative AI: A high school survey study of cheating behaviors before and after the release of ChatGPT. *Computers and Education: Artificial Intelligence*, *7*, 100253.

Ng, D. T. K., Chan, E. K. C., & Lo, C. K. (2025). Opportunities, challenges and school strategies for integrating generative AI in education. *Computers and Education: Artificial Intelligence*, *8*, 100373.

Wigfield, A., & Eccles, J. S. (2000). Expectancy–value theory of achievement motivation. *Contemporary Educational Psychology*, *25*(1), 68–81.

## Acknowledgments

We would like to thank Licenciado Hugo Gutiérrez Dávila, Secretary of Education of the State of Chihuahua, Mexico; Licenciado Humberto de las Casas, State Executive Director of COBACH (Colegio de Bachilleres); Licenciado José Luis García Rodríguez, Academic Dean of COBACH; and Licenciada Beatriz Acosta Salas for their invaluable support in distributing the survey. We are especially grateful to the students from public COBACH high schools across the state of Chihuahua for their participation in this study.

**Figure 1**
*AI Usage Patterns by Motivational Profiles*

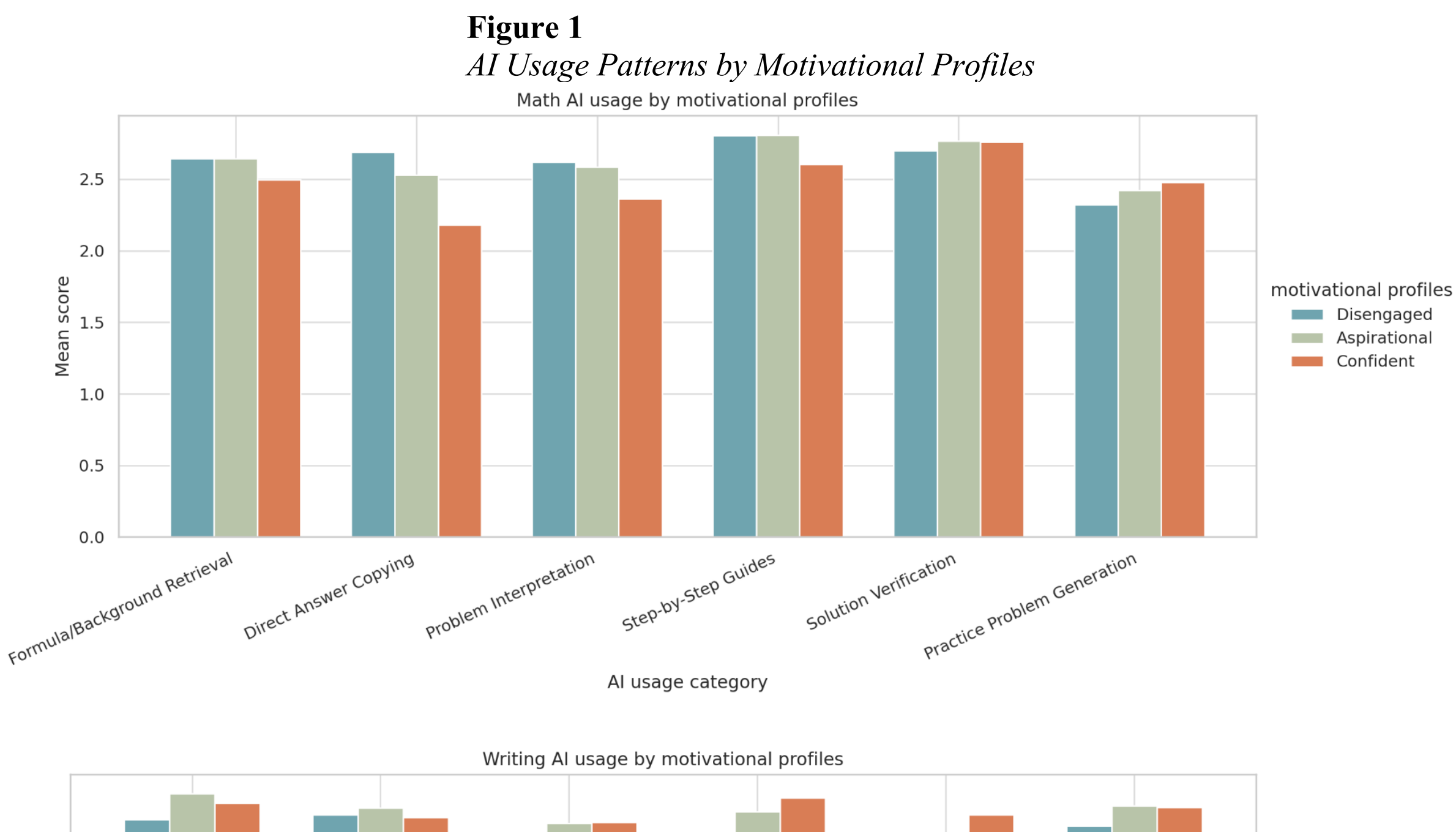

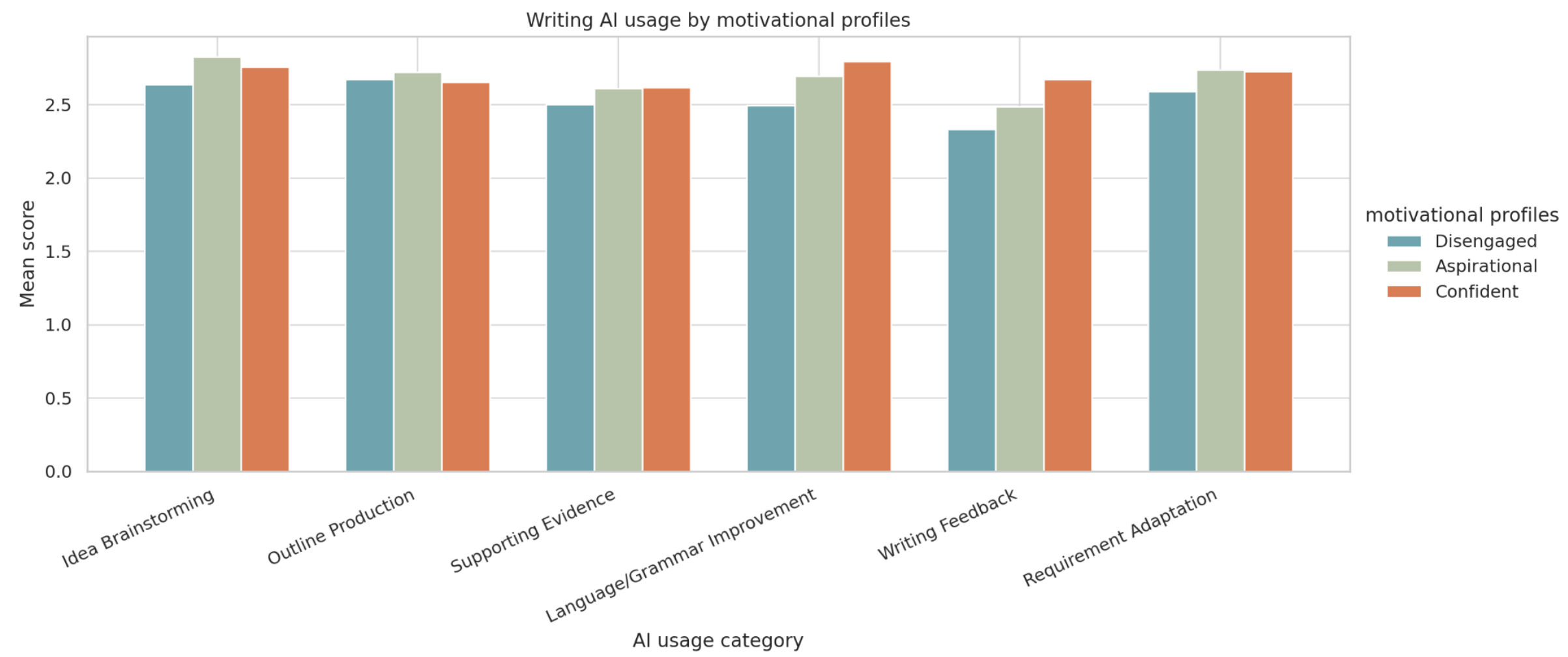